\documentclass[aps,prl,superscriptaddress,reprint]{revtex4-1}
\usepackage{amsmath} \usepackage{amsfonts}
\usepackage{amssymb} 
\usepackage{array}
\usepackage{graphicx} 
\usepackage{verbatim}

\begin{document}

\title{A ruthenium oxide thermometer for dilution refrigerators \\operating down to 5 mK}

\author{S.A. Myers}
\affiliation{Department of Physics and Astronomy, Purdue University, West Lafayette, IN 47907}
\author{H. Li}
\affiliation{Department of Physics and Astronomy, Purdue University, West Lafayette, IN 47907}
\author{G.A. Cs\'athy}
\affiliation{Department of Physics and Astronomy, Purdue University, West Lafayette, IN 47907}

\date{\today}
\begin{abstract}
At the lowest temperatures achieved in dilution refrigerators,
ruthenium oxide resistance thermometers often saturate and therefore lose their sensitivity.
In an effort to extend the range of such temperature sensors, we built a thermometer
which maintains sensitivity to 5~mK. A key component of this thermometer is an 
in situ radio frequency filter which is based on a modern rf absorption material.
We show that the use of such a filter is only effective when it is encased
in the same rf-tight enclosure as the ruthenium oxide sensor. Our design
delivers an attenuation level that is necessary to
mitigate the effects of parasitic heating of a fraction of pW present in our circuit.
Furthermore, we show that the likely origin of this parasitic heating is the
black body radiation present within the experimental space of the refrigerator.
We found that the equilibration time of the thermometer increases rapidly as the temperature is lowered;
below 5~mK this thermometer becomes impractical because of the prohibitively long equilibration times. 
\end{abstract}

\maketitle

\section{Introduction}
Dilution refrigerators are widely used in the study of condensed matter, nanoelectronics, and quantum technologies.
Modern instruments often cool to temperatures close to 5~mK. There is thus a need
to routinely measure such low temperatures. Among the large variety of thermometers
available for dilution refrigerators \cite{enss}, thick film ruthenium oxide resistors
(RuO$_2$) emerged as the most commonly used ones \cite{ruo-0,ruo-1,ruo-11,ruo-2,ruo-3,ruo-4,ruo-5,ruo-6,ruo-7,ruo-8,ruo-9}.
RuO$_2$ resistors are available as chips from electronics components vendors and from
companies specializing in low temperature thermometry.
Advantages of RuO$_2$ thermometers include wide availability, ease of use, small size, excellent
stability to thermal cycling, and moderate magnetoresistance.

Resistive RuO$_2$ thermometers, however, are not without their own limitations. In many cases they
become unusable typically below about 20~mK \cite{enss,ruo-0,ruo-1,ruo-11,ruo-2,ruo-3}. Indeed, 
at these temperatures the resistance versus 
temperature calibration curves of RuO$_2$ thermometers often saturate and 
these thermometers lose sensitivity  \cite{enss,ruo-0,ruo-1,ruo-11,ruo-2,ruo-3}. 
Furthermore, below 20~mK the resistance of RuO$_2$ thermometers is often not reproducible 
and may exhibit unexplained jumps in the resistance \cite{ruo-3}.
Such a behavior was attributed to a combination of two factors:
the gradual loss of thermal contact of electrons with the phonon bath as the temperature is lowered 
combined with parasitic radio frequency (rf) heating of electrons in the RuO$_2$ chip \cite{ruo-2}.
Therefore below about 20~mK one may not rely on these thermometers, and the calibration of these thermometers
at the lowest temperatures will likely be off when moved to a different dilution refrigerator.
The literature has two examples of RuO$_2$ based resistors that maintained
sensitivity to 6~mK \cite{ruo-8} and to 7.8~mK \cite{ruo-9}. 
These thermometers are, however, not based on commercial chips, but 
rather prepared from a RuO$_2$ paste \cite{ruo-8} or a proprietary ink \cite{ruo-9}. 
However, it is not clear if any material specific properties led to this improvement.
The loss of sensitivity of RuO$_2$ based resistors near the base temperature
of dilution refrigerators thus remains poorly understood, and a recipe for mitigating it
did not yet emerge. 
  
Parasitic rf heating plaguing RuO$_2$ thermometry at low temperatures
is also recognized to have detrimental effect in a variety of electronic systems.
Examples are single electron systems realized with metallic and semiconductor based quantum dots
\cite{devoret,nano1,nano2,nano3,nano}, sub-millikelvin Johnson noise 
thermometers \cite{saunders1,saunders2}, quantum Hall systems \cite{pan,book}, 
and superconducting and semiconductor quantum dot quantum processors \cite{qq}.
Parasitic heating in these systems is commonly controlled with rf filters built using
various absorbant materials. Examples of such filters are the ones based on 
epoxy impregnated metal powders \cite{pow1,pow2,pow3,pow4,pow5},
silver epoxy \cite{setup,ep}, thermocoax \cite{tc}, and magnetically loaded high loss dielectrics \cite{mag}.
However, to our knowledge, modern rf filtering technology was not used
for mitigating parasitic heating in RuO$_2$ thermometers.

In this report we set out to fabricate a RuO$_2$ resistance thermometer based on a commercial chip that is
capable of reaching temperatures as low as 5~mK. A key component of this thermometer is an in situ 
silver epoxy rf filter that takes advantage of several elements of modern rf filtering technology. Direct performance comparisons are made with unshielded RuO$_2$ resistance thermometers which are shown to saturate with temperature below 20~mK. This allows us to estimate the parasitic rf heating power picked up by our circuitry as well as determine the effective temperature of the black body radiation responsible for such heating. Lastly, we discuss the practicality of our newly designed RuO$_2$ resistance thermometer by analyzing how the thermal equilibration time scales with temperature.

\begin{figure}[t]
  \includegraphics[width=1\columnwidth]{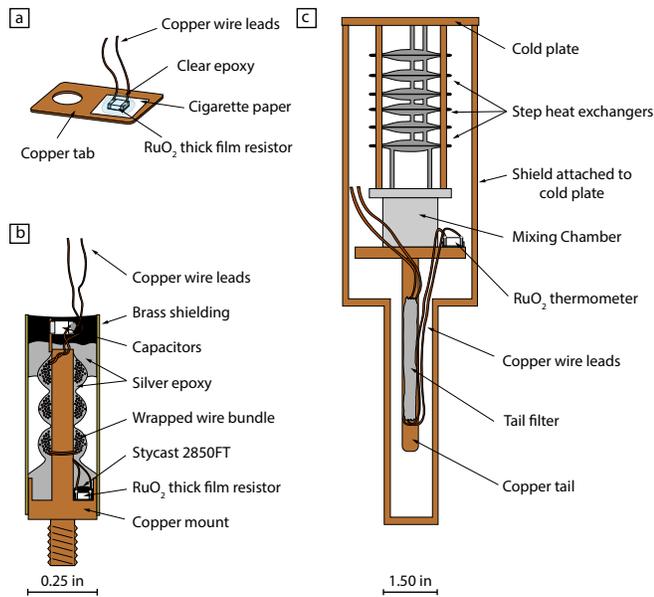} 
  \caption{Panel a: Schematic of thermometers Th-1, Th-2, and Th-3.
  Panel b: Schematic of Th-4. Panel c: Schematic of our measurement setup. The copper tail attached to the mixing chamber with a silver epoxy rf filter and the location of the RuO$_2$ 
  thermometer are clearly marked.  
\label{f1} }
\end{figure}

\begin{table}[b]
\centering
\begin{tabular}{ |c||c|c|c|  }
\hline
\multicolumn{4}{|c|}{RuO$_2$ Resistor Parameters} \\
\hline
Thermometer & Chip size & $R(\Omega)$ & Part number \\
\hline
Th-1  &  0805 &   738   &  RK73H2ALTD7380F   \\
Th-2  &  0805 &   300   &  RK73H2ATTD3000F   \\
Th-3  &  0402 &   560   &  RK73H1ETTP5600F   \\
Th-4  &  0402 &   560   &  RK73H1ETTP5600F   \\
\hline
\end{tabular}
\caption{Nominal parameters of the thick film RuO$_2$ resistor chips used for the thermometers discussed.
\label{t1}}
\end{table}

\subsection{Design of RuO$_2$ thermometers}
In this study we use several commonly available RuO$_2$ chips \cite{resistor}. Information on these chips can be found in Table.I. Thermometers
Th-1, Th-2, and Th-3 are built by attaching RuO$_2$ chips of different size
to a copper tab, and a schematic of the thermometer design
is shown in Fig.\ref{f1}a. Magnet wires are soldered onto the chip using indium,
and cigarette paper is used to electrically isolate the chips from the copper tab.
These chips are glued onto the copper tab face down, i.e. the conductive element and its protective insulation  
faces the cigarette paper and the copper tab.

A more elaborately designed thermometer Th-4 is shown in Fig.\ref{f1}b. Built into the rf-tight enclosure of Th-4 is a low-pass rf filter based on silver epoxy \cite{setup,ep}, which we refer to as the in situ filter. We chose silver epoxy as the absorbant because of its effectiveness
in absorbing rf power, its thermal conduction, and its non-magnetic property \cite{setup,ep}.
In order to enhance the effectiveness of the in situ filter, we impose the following constraints for Th-4:
1) the RuO$_2$ chip is mounted in an rf-tight enclosure 
2) the in situ rf filter is also located inside the rf-tight enclosure 
   and it is based on a twisted pair of magnet wires encased into a silver epoxy matrix 
3) wrapping of the twisted pair of wires that form the in situ filter
   is done in three well separated bundles and
4) two capacitors are used to channel rf waves to ground.
We thus used the best practices in rf filtering that are known to lower the
filter cut-off frequency, such as the use of capacitors \cite{pow5,ep} and of wire
bundles instead of evenly distributed windings \cite{ep}.
The novelty of our design is an rf-tight encapsulation of the RuO$_2$ chip, of the in situ filter, and
of the wires connecting the in situ filter and the RuO$_2$ chip.

We describe here the components and the assembly process for Th-4.
As shown in Fig.\ref{f1}b, the RuO$_2$ chip, the enclosure, and the in situ rf filter are supported by a threaded
copper mount machined out of a 1/4 inch diameter stock.
The region where the chip is mounted has a 2~mm deep
annular groove. A twisted pair of magnet wires of AWG 38 with heavy polyimide insulation
was soldered to the RuO$_2$ thick film resistor using indium. 
Afterwards, the RuO$_2$ chip was glued face down into the groove and epoxied using Stycast 2850FT. 
A small piece of cigarette paper was used to prevent an electrical short to the copper mount and
care was taken to minimize the amount of epoxy used. 
However, we made sure that the non-conductive epoxy covered all conductive parts, including the stripped magnet wire, 
contacts of the chip, and the indium solder.
The twisted wire pair of approximately 1.5 meters in length was then wrapped around the central copper column 
in three roughly equally spaced bundles. The silver epoxy \cite{silverepoxy} was applied between the layers, as the wires
were wrapped onto the post. 
We emphasize that care must be taken for the thermometer circuit not to short to ground:
we relied on the Stycast epoxy to electrically insulate any exposed metallic parts and on the heavy polyimide insulation 
of the wires to keep the silver epoxy from shorting to the magnet wires. 
We then soldered each magnet wire to a 1~nF capacitor \cite{capac}
connected to ground. These capacitors are mounted near the top of the thermometer.
A brass thin walled tube was slipped onto the copper mount; its bottom part was attached to the copper mount with
silver epoxy, while its top part was sealed with the same type of silver epoxy.
We note that when cooled, the small volume of trapped air within the brass tube will freeze. 
One may eliminate this open volume by filling it with silver epoxy. However, the large thermal mass of the epoxy may
negatively affect thermalization times of the thermometer.
Thus, we opted to use the minimum amount of silver epoxy required to produce an rf-tight enclosure.
Lastly, Stycast 2850FT was used to structurally hold the capacitors and magnet wires in place.

\subsection{Cryogenic setup for resistance thermometry measurement}
Our measurement setup has two rf filters: the room temperature and the tail filter.
The tail filter, shown in Fig.\ref{f1}c, is a silver epoxy filter located in a groove
along the copper tail that is fastened to the mixing chamber plate. We use this copper tail to thermalize samples 
exposed to magnetic fields generated by a superconductive solenoid. 
Along with rf attenuation, the silver epoxy used for the tail filter
provides structural support and thermally anchors the electrical leads
that run along the tail. Details of these filters 
can be found in Ref.\cite{setup}. As shown in Fig.\ref{f1}c,
the RuO$_2$ thermometers are mounted on the mixing chamber plate and connect to the bottom end of the tail filter. The measurement wires exiting through the top of the tail filter
are heatsunk at the mixing chamber and connect to room temperature via 0.1~mm constantan wires 
provided with our dilution refrigerator by the manufacturer, Oxford Instruments. 
These constantan wires have a resistance near 210~$\Omega$ and are heatsunk at each cooling stage of the instrument. 

The four thermometers were calibrated against a carbon thermometer described in Ref.\cite{carbon}. 
This carbon thermometer was shown to thermalize to low temperatures and was
calibrated against a He-3 quartz tuning fork viscometer \cite{setup}. 
In comparison to the carbon thermometer, we expect that our RuO$_2$ thermometers 
will tolerate much reduced heating levels \cite{ruo-9,setup}.
Four-wire resistance measurements were performed using a Lakeshore 370 AC resistance bridge 
with a model 3716 scanner front end.
As shown later, an excitation level of 100~pA was chosen at the lowest temperatures to avoid self-heating of the RuO$_2$ thermometers.


\begin{figure}[t]
  \includegraphics[width=1\columnwidth]{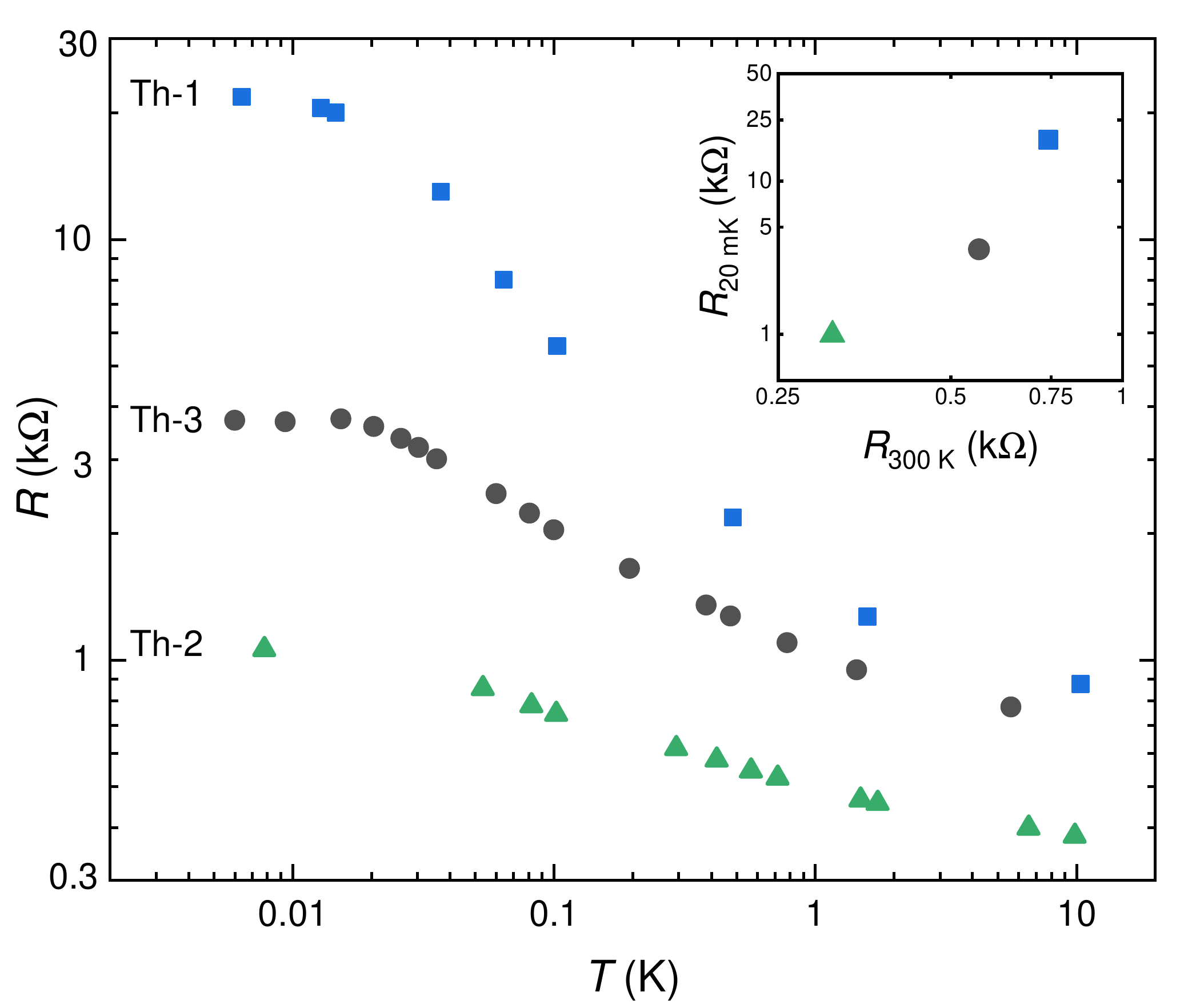} 
  \caption{Resistance calibration for various RuO$_2$ thermometers. 
  Thermometers Th-1, Th-2, and Th-3 have a similar design; the RuO$_2$ chip
  are attached to a copper tab and cigarette paper is used to prevent shorts to ground.
  The inset shows the relationship of $R_{300 K}$ and $R_{20 mK}$, i.e. resistances measured at
  room temperature and at 20~mK.
\label{f2} }
\end{figure}

\subsection{Calibration curves for RuO$_2$ thermometers Th-1, Th-2, and Th-3}
In Fig.\ref{f2} we show the calibration of the three RuO$_2$ thermometers affixed to copper tabs.
As expected, the resistance of these thermometers increases with a decreasing temperature.
As shown in the inset of Fig.\ref{f2}, larger room temperature values result in larger 
low temperature values. In addition, for Th-1 and Th-3
we observe a decrease in their sensitivity starting at 30~mK, and there is a clear saturation
of the resistance below 20~mK. 
This behavior is similar to that reported in Refs.\cite{ruo-0,ruo-1,ruo-11,ruo-2,ruo-3}, and we associate
this saturation of the calibration curve with the presence of unwanted parasitic rf heating. 
We recall that the circuit
for measuring these thermometers contains two rf filters: one at room temperature and another
silver epoxy based filter affixed to the tail.  
We thus find that a setup with these two rf filters does not result in an effective electronic
thermalization within the RuO$_2$ chips of the three thermometers based on copper tabs.

\subsection{Low temperature performance of Th-4 with an in situ filter}

\begin{figure}[t]
  \includegraphics[width=1\columnwidth]{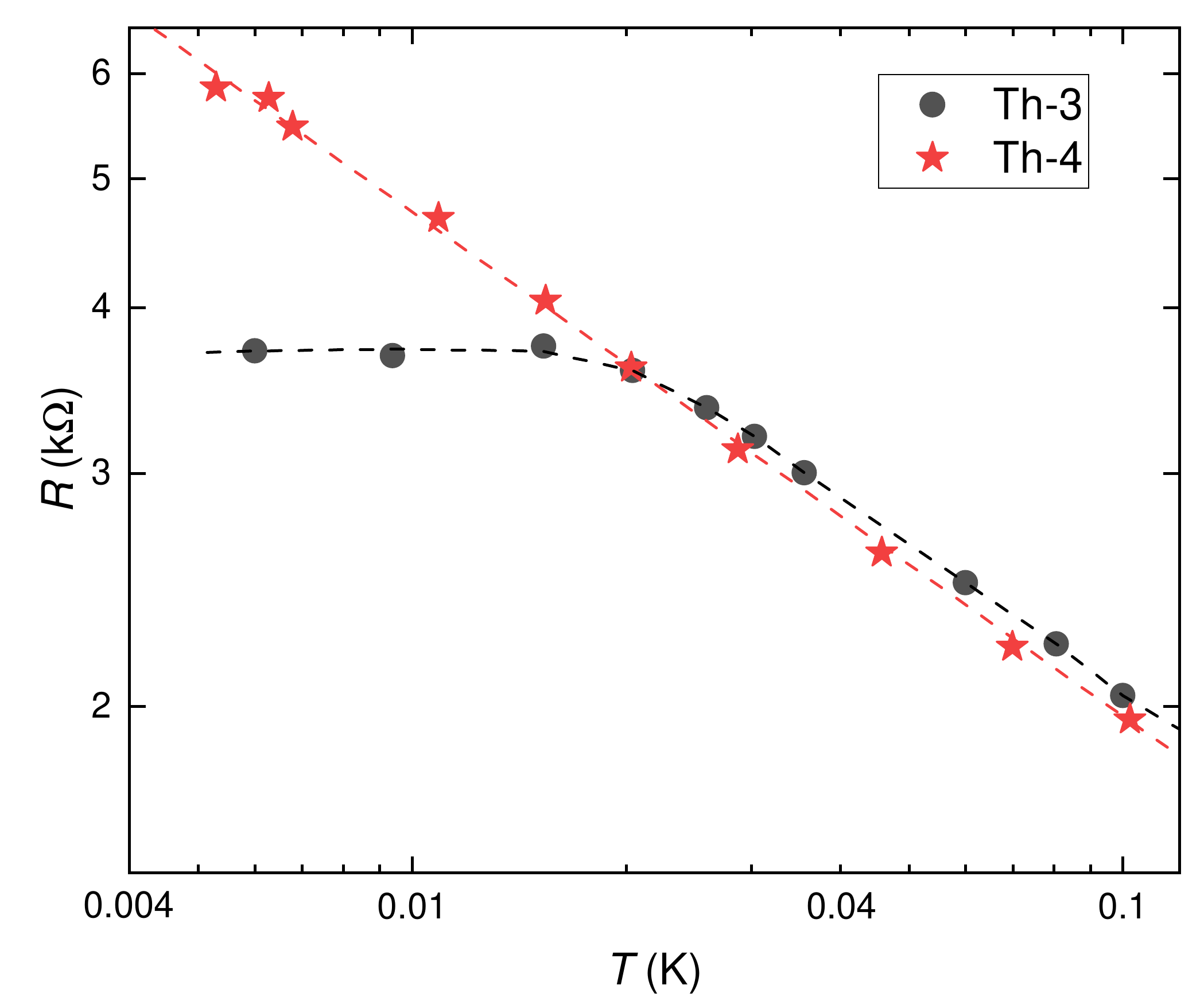} 
  \caption{A comparison of the low temperature calibration of Th-3 and Th-4. These thermometers
  are built using RuO$_2$ chips of identical size and nominal resistance, but Th-4 has an in situ 
  rf filter and is sealed in an rf-tight enclosure. Dashed lines are guides for the eye.
  The calbration curve of Th-3 saturates below about 20~mK. In stark contrast, Th-4
  maintains its sensitivity to the lowest temperatures.
\label{f3} }
\end{figure}

In Fig.\ref{f3} we compare the resistance versus temperature calibration curves of Th-3 and Th-4. 
Even though built using RuO$_2$ chips from the same batch, the behavior of these two thermometers is quite different.
Above 20~mK, both thermometers have a similar trend. Small changes in the calibration,
such as the ones shown in Fig.\ref{f3}, are expected when using chips taken from the same batch.
However, the most striking difference between the two thermometers
is in their low temperature behavior. Below about 20~mK, the resistance
of Th-4 continues to rise monotonically as the temperature is lowered. 
In fact the resistance as function of temperature for Th-4 maintains the same power 
law $R \propto T^{-0.38}$ between 0.1~K and 5~mK, indicating
therefore no change in the conduction mechanism over this temperature range.
Thermometer Th-4 can thus be used to measure temperature at least down to 5~mK,
extending therefore the useful range for a RuO$_2$ thermometer to the full range of
dilution refrigerator temperatures. Maintaining sensitivity down to 5~mK in Th-4 indicates the successful mitigation of the  unwanted parasitic rf heat in this thermometer. We attribute the removal of the parasitic rf heat to the combination
of the rf-tight enclosure and the effects of the in situ rf filter.

\subsection{Self-heating analysis}
In the most general case, the power dissipated through a RuO$_2$ chip is
$P=I^2 R+P_{paras}$. Here $I$ is the current excitation, $I^2 R$ the Joule heating of this
excitation, and $P_{paras}$ the parasitic rf heating. 
However, the in situ filter used in the construction of Th-4 effectively dissipates the parasitic rf heating.
Thus for Th-4 the parasitic rf heating is assumed to be negligible $P_{paras} \approx 0$. 
The non-saturating resistance of Th-4 at the lowest temperatures, shown in Fig.\ref{f3}, 
supports the validity of this assumption.
\begin{figure}[htb!]
  \includegraphics[width=1\columnwidth]{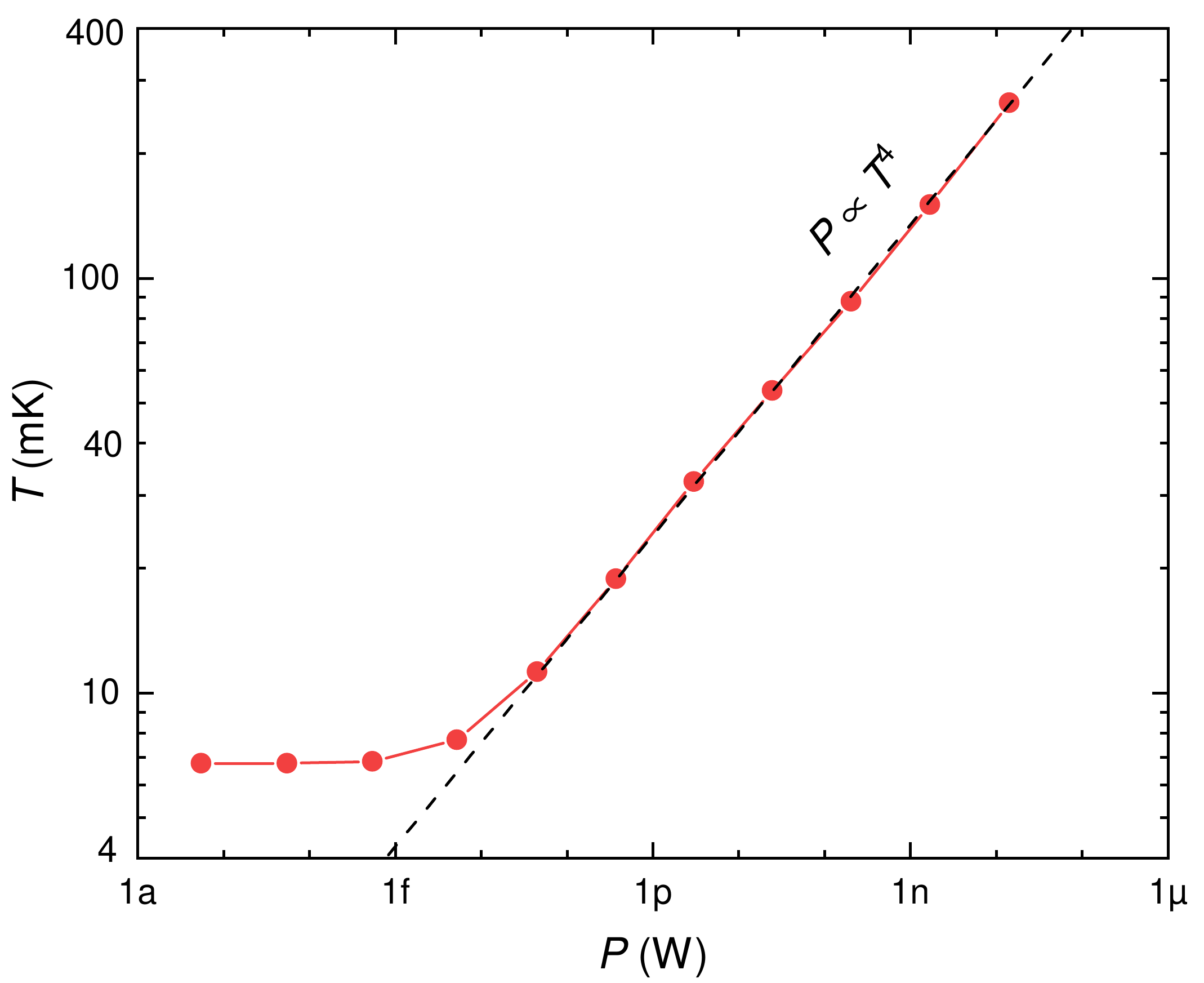} 
  \caption{Self-heating curve $T$ versus the Joule excitation power $P$ of the thermometer Th-4
   built with an in situ rf filter. 
  The mixing chamber during this measurement was held at $T=6.7$~mK. The temperature reading
  of this thermometer starts to rise at an excitation level of about 1 fW. The dashed line is $P \propto T^4$.
  \label{f4} }
\end{figure}
By performing a so called self-heating experiment, we establish the upper bound on the excitation levels which will not overheat this thermometer.
For this measurement the mixing chamber was held fixed at $T=6.7$~mK
and $I$ was successively increased. With an increasing excitation level, Joule heating causes the 
electronic temperature to rise, hence the resistance decreases.
The observed drop in resistance was converted to an increase in temperature using the calibration
curve shown on Fig.\ref{f3}. The resulting self-heating curve is shown in Fig.\ref{f4}.
We found that near the base temperature the electron temperature 
in the conductive channel of the RuO$_2$ chip starts
decoupling from the phonon bath temperature at around 1~fW of heating power. 
This value is comparable with measured values at 8~mK \cite{ruo-9,setup}. Such power levels require excitation currents of about $I=100$~pA. 
Hence such low excitation levels were used for calibration at the low end of the temperature scale.

The self-heating measurement also allows us to characterize the rf environment of the thermometers.
In contrast to Th-4, Th-3 lacks the in situ rf filter and therefore, is not thermalized to the base temperature due to parasitic rf heating.
We recall that the resistance of Th-3 saturates near $T=20$~mK.
According to the self-heating data shown in Fig.\ref{f4}, 
a Joule heating power of 0.45~pW is needed to heat Th-4 to 20~mK when the mixing chamber is kept cold.
Thus, we surmise the same power level $P_{paras}=0.45$~pW is present in our circuit and causes the saturation of the resistance of Th-3.

Since Th-4 thermalizes to 5~mK, its in situ filter is effective in attenuating the 0.45~pW parasitic power
to below 1~fW. We therefore estimate the attenuation level required for the in situ filter in order to effectively 
cool our RuO$_2$ chip to 5~mK to be $\approx 10 \log 0.45$~pW$/1$~fW$=27$~dB, or nearly
a factor of 1000. Such a figure is reasonable since
a filter built of similar elements had an attenuation exceeding this value in the 2~MHz to 10~GHz frequency
band \cite{ep}. Nonetheless, one must keep in mind that those attenuation measurements
were done in a 50~$\Omega$ environment.
In the present case, a direct measurement of the spectral distribution of the attenuation in a RuO$_2$ chip
is not readily possible since changes with temperature of the RuO$_2$ load resistance will cause
a change in the frequency dependent attenuation. Evidence for such an effect
can be found in a Johnson noise thermometry experiment in which
parasitic heating led to the saturation of the electronic temperature that was dependent on the 
load resistance \cite{saunders2}. We think that in contrast to the broad band nature of
microwave circuits built with coaxial lines and with $50$~$\Omega$ sources and loads, 
in our circuit the power is effectively coupled into Th-3 through 
a set of discrete frequencies. Therefore in our circuit
a bolometric measurement of the power through the earlier described self-heating measurement
is more meaningful.

\subsection{Effective temperature of black body radiation}
In order to identify the origin of the 0.45~pW parasitic power the thermometer Th-3 experiences,
it is important to recognize that rf radiation may couple to 
this thermometer in two distinct ways. First, rf heating may originate from
outside the dilution refrigerator; rf waves from our lab
may travel along the measurement wires toward the RuO$_2$ chip.
Second, rf heating may originate from inside the dilution refrigerator, as a black body radiation
present in the environment of the RuO$_2$ chip. As seen in Fig.\ref{f1}c, a length of about 0.3 meters of the measurement wires
leaving the bottom of the tail filter and connecting to the RuO$_2$ chip on the mixing chamber plate
is not shielded and is thus exposed to the black body radiation present in this region.
The most striking difference in the construction of Th-3 and Th-4
is that the former is directly exposed to this black body radiation, whereas the latter is not. 
We thus think that a major contributor to the saturation of the low temperature calibration of Th-3 is 
this black body radiation. Assuming that the measurement wires are fully thermalized by the heat sinks and tail filter and assuming an ideal emisittivity, we can provide a crude estimate
for $\widetilde{T}$, the effective temperature of a black body radiating the $P=0.45$~pW power into 
our thermometer circuit. 
From the Stefan-Boltzmann law $P=\sigma A \widetilde{T}^4$ we estimate $\widetilde{T} \approx 0.4$~K. 
Here $A$ is the area of the exposed wires that run from the bottom of the tail filter to the thermometer. 
$\widetilde{T}$ is significantly larger than the 90~mK temperature of the cold plate
and the radiation shield attached to it.  Black body radiation from the liquid helium bath of our wet refrigerator
penetrating through openings of the cold plate may explain such an effective temperature $\widetilde{T}$.
In typical instruments, such as ours, the cold plate and its shield do not form an rf-tight enclosure
as there are through holes and openings on the cold plate near the wiring, for example.
Our results highlight therefore the necessity of blocking such a radiation from entering the channel of the RuO$_2$
chip. Moreover, the four constraints put forth in the design of Th-4 can also be used
to effectively cool any electronic device to millikelvin temperatures or below.

\subsection{Equilibration time of Th-4}

\begin{figure}[htb!]
  \includegraphics[width=1\columnwidth]{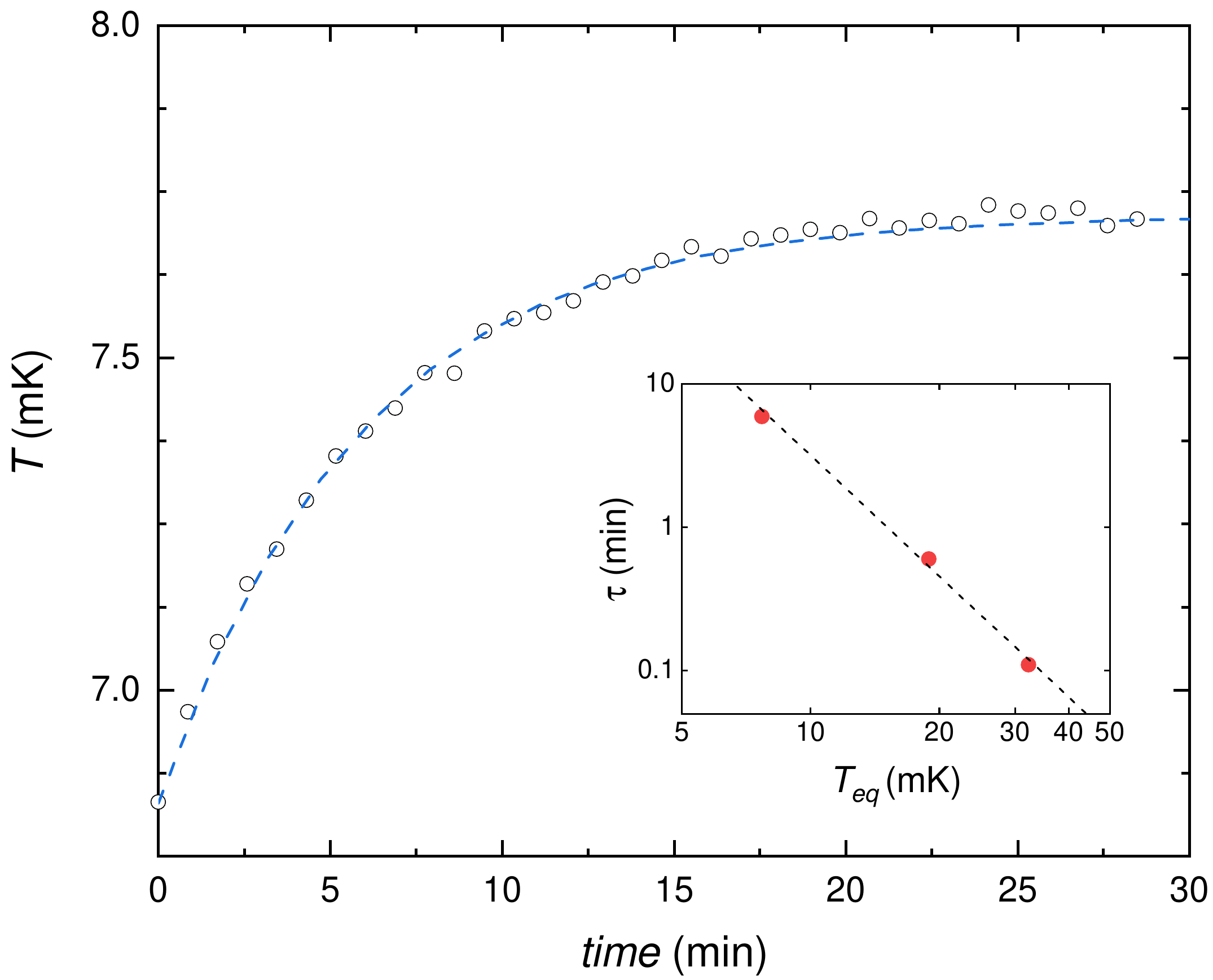} 
  \caption{Time evolution reading of Th-4 when its excitation is increased from 316~pA to 1~nA.
  Here the mixing chamber temperature is held at $6.7$~mK. 
  The dashed line is an exponential fit from which the time constant $\tau$ is
  extracted. The inset shows the dependence of the time constant $\tau$ on the
  equilibration temperature $T_{eq}$ of the conduction channel of the thermometer. The dashed line is a power law fit of the time constants $\tau \propto T_{eq}^{-2.8}$.
\label{f5} }
\end{figure}

While the improved low temperature resistive behavior of Th-4 is evident when compared against a typical RuO$_2$ thermometer,
such as Th-3, there remains another critical aspect for practicality purposes: the time required for the thermometer to equilibrate. 
It is known that the thermal response of any type of resistive thermometer is strongly temperature dependent
in the millikelvin temperature range.
We quantify trends in the equilibration time by analyzing the transient behavior 
after injecting additional power into the thermometer.
A representative response is shown in Fig.\ref{f5}. 
The transient behavior is exponential to a good degree.
Time constants $\tau$ are extracted when the mixing chamber temperature is held at a fixed value 
and the excitation current is successively increased from 316~pA to 1~nA, 3.16~nA, and 10~nA. Shown in the inset of Fig.\ref{f5} is the dependence $\tau$ on the equilibration temperature $T_{eq}$, the temperature at which transient behavior subsides. We find that for temperatures below 40~mK, $\tau$ increases rapidly according to 
the power law $\tau \propto T_{eq}^{-2.8}$. Thus, while Th-4 remains sensitive at temperatures below 20~mK, 
the time required for the thermometer to stabilize becomes increasingly longer with lower temperatures.
Even though Th-4 may be used down to 5~mK, we expect that its response time at significantly lower temperatures
becomes prohibitively long to a point at which RuO$_2$ resistive thermometry becomes impractical.

\subsection{Conclusions}
To summarize, we designed a RuO$_2$ thick film resistor based thermometer which follows the phonon bath temperature 
down to 5~mK. 
We achieved such a thermalization by attenuating parasitic radio frequency waves traveling along the measurement
wires using an in situ silver epoxy filter and by using an rf-tight enclosure of the RuO$_2$ chip, the in situ
filter, and wiring connecting these elements.
Furthermore, we think that the parasitic radio frequency heating originated not from the room temperature parts of our circuit but instead from the black body radiation present at the location of the thermometer.
Thus, our findings highlight that rf shielding within the experimental space of both the electrical components and measurement wires is likely necessary for many devices to reach low millikelvin temperatures and below.
From self-heating curves we found that the thermometer's in situ rf filter effectively dissipates the 0.45~pW 
of radio frequency heating power.
Lastly, analysis of the self-heating transient behavior shows an undesirable rapid increase in the equilibration 
times with decreasing temperature. Thus, the striking improvement in sensitivity comes at a cost of increased 
thermal response times, seriously limiting the use of the thermometer below 5~mK.

\subsection*{Acknowledgments}

This work was supported by the US DOE, Office of Basic
Energy Sciences under the award DE-SC0006671.


\begin{thebibliography}{l}

\bibitem{enss} C. Enss and S. Hunklinger,  \textit{Low-Temperature Physics}, Springer, Berlin (2005).

\bibitem{ruo-0} H. DOI, Y. Narahara, Y. Oda, and H. Nagano, \textit{New resistance thermometer with small magnetic field dependence for low temperature measurements}, Proceedings of the 17th International Conference on Low Temperature Physics LT-17, 405 (1984).

\bibitem{ruo-1} Q. Li, C. Watson, R. Goodrich, D. Haase, and H. Lukefahr, \textit{Thick film chip resistors for use as low temperature thermometers}, Cryogenics \textbf{2}, 467 (1986).
https://doi.org/10.1016/0011-2275(86)90095-0

\bibitem{ruo-11} W. Schoepe, \textit{Conduction mechanism in granular Ru0$_2$-based thick-film resistors}, Physica B \textbf{165}/\textbf{166}, 299 (1990).
10.1016/s0921-4526(90)80999-y

\bibitem{ruo-2} M. Watanabe, M. Morishita, and Y. Ootuka, \textit{Magnetoresistance of RuO$_2$-based resistance thermometers below 0.3 K}, Cryogenics \textbf{41} 143 (2001).
https://doi.org/10.1016/s0011-2275(01)00066-2


\bibitem{ruo-3} P. Jorba, \textit{Calibration of RuO$_2$ commercially available resistors for temperature sensing at
dilution refirerator temperatures}, TP IV Report supervised by H. Ronnow and J. Piatek,
l'Ecole Polytechnique F\'ed\'erale de Lausanne (2013).  


\bibitem{ruo-4} A. Briggs, \textit{Characterization of some chip resistors at low temperatures}, Cryogenics \textbf{31}, 932 (1991).
https://doi.org/10.1016/0011-2275(91)90216-j

\bibitem{ruo-5} K. Uhlig, \textit{Magnetoresistance of thick-film chip resistors at millikelvin temperatures}, Cryogenics \textbf{35}, 525 (1995).
https://doi.org/10.1016/0011-2275(95)98220-u


\bibitem{ruo-6} R. Goodrich, D. Hall, E. Palm, and T. Murphy, \textit{Magnetoresistance below 1K and temperature cycling of ruthenium oxide–bismuth ruthenate cryogenic thermometers}, Cryogenics \textbf{38}, 221 (1998).
https://doi.org/10.1016/s0011-2275(97)00100-8

\bibitem{ruo-7}  C.J. Yeager, S.S. Courts, and W.E. Davenport, \textit{Thermal resistance of cryogenic thermometers at ultra-low temperatures}, AIP Conference Proceedings \textbf{613}, 1644 (2002).  
https://doi.org/10.1063/1.1472201

\bibitem{ruo-8} R. Willekers, F. Mathu, H. Meijer, and H. Postma, \textit{Thick film thermometers with predictable R-T characteristics and very low magnetoresistance below 1 K}, Cryogenics \textbf{30}, 351 (1990).
https://doi.org/10.1016/0011-2275(90)90315-4

\bibitem{ruo-9} S.S. Courts, and J.K. Krause, \textit{A commercial ruthenium oxide thermometer for use to 20 millikelvin}, AIP Conference Proceedings \textbf{985}, 947 (2008). 
https://doi.org/10.1063/1.2908694

\bibitem{devoret} 
D. Vion, P.F. Orfila, P. Joyez, D. Esteve, and M.H. Devoret, \textit{Miniature electrical filters for single-electron devices},
J. Appl. Phys. \textbf{77}, 2519 (1995).
https://doi.org/10.1063/1.358781

\bibitem{nano1}  L. Casparis, M. Meschke, D. Maradan, A.C. Clark, C.P. Scheller, K.K. 
Schwarzw\"{a}lder, J.P. Pekola, and D.M. Zumb\"{u}hl, \textit{Metallic Coulomb blockade thermometry down to 10 mK and below}, Rev. Sci. Instrum. \textbf{83}, 083903 (2012).
https://doi.org/10.1063/1.4744944

\bibitem{nano2} D. Maradan, L. Casparis, T.-M. Liu, D.E.F. Biesinger, C.P. Scheller, D.M. Zumb\"{u}hl, 
J.D. Zimmerman, and A.C. Gossard, \textit{GaAs quantum dot thermometry using direct transport and charge sensing}, J. Low Temp. Phys. \textbf{175}, 784 (2014).
https://doi.org/10.1007/s10909-014-1169-6

\bibitem{nano3} M. Sarsby, N. Yurttag\"{u}l, and A. Geresdi, \textit{500 microkelvin nanoelectronics}, Nat. Commun. \textbf{11}, 1492 (2020).
https://doi.org/10.1038/s41467-020-15201-3

\bibitem{nano} A.T. Jones, C.P. Scheller, 
J.R. Prance, Y.B. Kalyoncu, D.M. Zumb\"{u}hl, and R.P. Haley, \textit{Progress in cooling nanoelectronic devices to ultra-low temperatures}, J. Low Temp. Phys. \textbf{201}, 772 (2020).
https://doi.org/10.1007/s10909-020-02472-9


\bibitem{saunders1} C.P. Lusher, J. Li, V.A. Maidanov, M.E. Digby, H. Dyball, A. Casey, J. Ny\'eki, V.V. Dmitriev,
B.P. Cowan, and J. Saunders, \textit{Current sensing noise thermometry using a low $T_\text{c}$ DC SQUID preamplifier}, Meas. Sci. Technol. \textbf{12}, 1 (2001).
https://doi.org/10.1088/0957-0233/12/1/301

\bibitem{saunders2} A. Casey, F. Arnold, L.V. Levitin, C.P. Lusher, J. Ny\'eki, J. Saunders,
A. Shibahara, H. van der Vliet, B. Yager, D. Drung, Th. Schurig, G. Batey, M.N. Cuthbert, and A.J. Matthews, \textit{Current sensing noise thermometry: A fast practical solution to low temperature measurements},
J. Low Temp. Phys. \textbf{175}, 764 (2014).
https://doi.org/10.1007/s10909-014-1147-z

\bibitem{pan} W. Pan, J.-S. Xia, V. Shvarts, D.E. Adams, H.L. Stormer, D.C. Tsui, L.N. Pfeiffer, 
K.W. Baldwin, and K.W. West, \textit{Exact quantization of the even-denominator fractional quantum Hall state at $\nu=5/2$ Landau level filling factor}, Phys. Rev. Lett. \textbf{83}, 3530 (1999).
https://doi.org/10.1103/PhysRevLett.83.3530

\bibitem{book} G.A. Cs\'athy, chapter 5 in \textit{Fractional Quantum Hall Effects: New Developments},
edited by B.I. Halerin and J.K. Jain, (World Scientific, 2020).

\bibitem{qq}
N.P. de Leon, K.M. Itoh, D. Kim, K.K. Mehta, T.E. Northup, H. Paik,
B.S. Palmer, N. Samarth, S. Sangtawesin, and D.W. Steuerman, \textit{Materials challenges and opportunities for quantum computing hardware},
Science \textbf{372}, 6539 (2021).
https://doi.org/10.1126/science.abb2823

\bibitem{pow1} J.M. Martinis, M.H. Devoret, and J. Clarke, 
\textit{Experimental tests for the quantum behaviour of a macroscopic degree of freedom: The phase difference across a Josephson junction},
Phys. Rev. B \textbf{35}, 4682 (1987).
https://doi.org/10.1103/PhysRevB.35.4682

\bibitem{pow2} A. Fukushima, A. Sato, A. Iwasa, Y. Nakamura, T. Komatsuzaki, and Y. Sakamoto. \textit{Attenuation of microwave filters for single-electron tunneling experiments} 
IEEE Trans. Instrum. Meas. \textbf{46}, 289 (1997).
https://doi.org/10.1109/19.571834

\bibitem{pow3} K. Bladh, D. Gunnarsson, E. H\"{u}rfeld, S. Devi, C. Kristoffersson, B. Sm$\mathring{\text{a}}$lander, 
S. Pehrson, T. Claeson, P. Delsing, and M. Taslakov, \textit{Comparison of cryogenic filters for use in single electronics experiments}, Rev. Sci. Instrum. \textbf {74}, 1323 (2003).
https://doi.org/10.1063/1.1540721

\bibitem{pow4} F.P. Milliken, J.R. Rozen, G.A. Keefe, and R.H. Koch, \textit{50~$\Omega$ characteristic impedance low-pass metal powder filters}, 
Rev. Sci. Instrum \textbf{78}, 024701 (2007).
https://doi.org/10.1063/1.2431770

\bibitem{pow5} A. Lukashenko, and A.V. Ustinov, \textit{Improved powder filters for qubit measurements}, Rev. Sci. Instrum. \textbf{79}, 014701 (2008).
https://doi.org/10.1063/1.2827515

\bibitem{setup} N. Samkharadze, A. Kumar, M.J. Manfra, L.N. Pfeiffer, K.W. West, and G.A. Cs\'athy, \textit{Integrated electronic transport and thermometry at milliKelvin temperatures and in strong magnetic fields} 
Rev. Sci. Instrum. \textbf {82}, 053902 (2011).
https://doi.org/10.1063/1.3586766

\bibitem{ep} C.P. Scheller, S. Heizmann, K. Bedner, D. Giss, M. Meschke, D.M. Zumb\"{u}hl, J.D. Zimmerman, and A.C. Gossard, \textit{Silver-epoxy microwave filters and thermalizers for millikelvin experiments}, App. Phys. Lett. \textbf{104}, 21106 (2014).
https://doi.org/10.1063/1.4880099

\bibitem{tc} A.B. Zorin, \textit{The thermocoax cable as the microwave frequency filter for single electron circuits},  Rev. Sci. Instrum. \textbf{66}, 4296 (1995).
https://doi.org/10.1063/1.1145385

\bibitem{mag} D.F. Santavicca and D.E. Prober, \textit{Impedance-matched low-pass stripline filters}, Meas. Sci. Technol. \textbf{19}, 087001 (2008).
https://doi.org/10.1088/0957-0233/19/8/087001

\bibitem{resistor} RK73H series from KOA Speer Electronics Inc., purchased from Digikey or Mouser

\bibitem{silverepoxy} MG Chemicals, Adhesive 8331D-A/B

\bibitem{capac} GRM1885C1H102JA01J   from Murata 

\bibitem{carbon} N. Samkharadze, A. Kumar, and G. A. Cs\'athy, \textit{A new type of carbon resistance thermometer with excellent thermal contact at millikelvin temperatures}, J. Low Temp. Phys. \textbf{160}, 246 (2010).
https://doi.org/10.1007/s10909-010-0192-5




\end{thebibliography}
\end{document}